# Large-Area Nanopatterned Graphene For Ultrasensitive Gas Sensing

Alberto Cagliani, David Mackenzie, Lisa Katharina Tschammer, Filippo Pizzocchero, Kristoffer Almdal and Peter Bøggild

DTU Nanotech – Center for Nanostructured Graphene (CNG) and Department of Micro- and Nanotechnology, Technical University of Denmark, Building 345 Ørsteds Plads, 2800 Kgs. Lyngby, Denmark

**ABSTRACT**

Chemical vapor deposited graphene is nanopatterned by a spherical block-copolymer etch mask. The use of spherical rather than cylindrical block copolymers allows homogeneous patterning of cm-scale areas without any substrate surface treatment. Raman spectroscopy was used to study the controlled generation of point defects in the graphene lattice with increasing etching time, confirming that alongside the nanomesh patterning, the nanopatterned CVD graphene presents a high defect density between the mesh holes. The nanopatterned samples showed sensitivities for $NO_2$ of more than one order of magnitude higher than for non-patterned graphene. $NO_2$ concentrations as low as 300 ppt were detected with an ultimate detection limit of tens of ppt. This is so far the smallest value reported for not UV illuminated graphene chemiresistive $NO_2$ gas sensors. The drastic improvement in the gas sensitivity is believed to be due to the high adsorption site density, thanks to the combination of edge sites and point defect sites. This work opens the possibility of large area fabrication of nanopatterned graphene with extreme density of adsorption sites for sensing applications.

# 1. Introduction

After the seminal discovery of microexfoliated graphene by Geim and Novoselov in 2004, graphene has been suggested as a candidate for chemical sensing with the possibility of detecting individual molecules and charges [1,2]. Due to the extreme surface-to-volume ratio, low intrinsic noise and tenability of the doping level, detected concentrations from hundreds ppm down to ppb have been reached for $NO_2$ and $NH_3$, using exfoliated graphene and reduced graphene oxide [2,3,4]. The availability of large-area graphene by chemical vapor deposition (CVD), has made mass fabricated graphene based sensor devices possible [5,6]. The simplest detection scheme is used in chemiresistive sensors, where adsorbed molecules lead to a change in the Fermi level of graphene. In the case of the oxidizing and highly toxic gas $NO_2$, the charge transfer occurs from graphene to the $NO_2$ open shell molecules thus increasing the hole concentration in the graphene sheet [7]. Unmodified CVD graphene based gas sensors and epitaxially grown graphene based sensors able to detect concentrations in the hundreds ppm down to 0.1 ppm range have been demonstrated [8,9] and their cross sensitivity with common interfering gases investigated [9-11]. The p-doping from $NO_2$ has been also confirmed by analysis of the transfer characteristic of field effect graphene transistors upon $NO_2$ adsorption [10,12].

However, such detected concentrations should be compared with the international standards for air quality. For example, the National Ambient Air Quality Standard (NAAQS) sets a limit exposure of 0.053 ppm (annual average) for $NO_2$ [13]. Moreover, Hesterberg and coauthors in their comprehensive analysis of the human clinical studies for $NO_2$ exposure, recommend a maximum concentration of 0.2 ppm [13]. These considerations imply that the relevant range of $NO_2$ ambient concentrations is in the sub-0.1 ppm in order to prevent any health hazard. The $NO_2$ minimum detected concentrations shown so far by CVD graphene based gas sensors are not sufficient to monitor $NO_2$ concentration in this range. In order to improve the gas sensitivity and the detection limits, different approaches have been explored, even if just few of them are immediately scalable. An increase in the sensitivity has been achieved thanks to the introduction of structural defects, which provide extra adsorption sites for $NO_2$ molecules. While different types of defects do not bind $NO_2$ equally well, defects will generally lead to a stronger gas response than pristine graphene, which has no dangling bonds in the basal plane [14,15]. Hajati and coauthors have experimentally demonstrated a three times increase in $NO_2$ sensitivity after ion bombardment and have theoretically shown that Stone-Wales structural defects are the most effective in binding $NO_2$ [16]. It has been also demonstrated that treatment of CVD graphene by ozone can double the sensitivity with respect to pristine CVD graphene by creating defects saturated with oxygen groups, reaching a detected concentration of 0.2 ppm with an extrapolated detection limit of 1.3 ppb [17]. Recently, the introduction of a graphene nanomesh showed a promising chemiresistive performance [18]. Paul and coauthors fabricated nanomeshed CVD graphene using colloidal (microsphere) lithography and thereby obtained an improvement of approximately a factor of 4.5 (at 2 ppm) compared to unpatterned CVD graphene. The improvement is attributed to the large edge length per unit area of the nanomesh, which provides extra edge adsorption sites for the gas molecules.

Most attempts at fabricating nanomeshed graphene have so far been motivated by the possibility of introducing an energy band gap [19,20,21]. Several self-aligning or bottom-up techniques have been used to fabricate graphene nanomesh, of which block copolymer (BCP) lithography is the most widely used [19,20,22,23]. In terms of gas sensing, this patterning technique offers the possibility of achieving very high edge length per unit area, which would increase the number of edge adsorption sites for the analyte molecules. All the works on BCP nanopatterning of graphene nanomesh used vertically aligned cylindrical BCPs, which require the surface energy to be accurately tuned in order not to favor either of the two polymeric constituents, as this will lead to the collapse of the vertical alignment of the cylinders upon annealing [24]. While a combination of long-range order and very small feature sizes can be achieved with cylindrical BCPs, the method requires complicated processing, with the largest nanopatterned graphene areas shown in literature limited to ~1 $cm^2$ [23].

So far, the different methods to enhance the chemiresponse of graphene based gas sensors have been investigated separately and without testing the gas sensors in the range relevant to prevent hazard to human health (sub 0.1 ppm). To the best of our knowledge only two published works have achieved enough sensitivity to test their graphene based gas sensors in the sub 0.1 ppm range. A drastic improvement of the gas sensitivity was reported for CVD graphene exposed in-situ to UV light [25]. However, in another study carbon nanotubes were shown to lose the sensor performance upon UV irradiation after few hours and the role of UV light during the gas sensing is still not understood [25,26]. Moreover, continuous UV illumination would strongly limit the portability, increase the cost and power consumption of such a sensor. Detection in the concentration range of interest was also reported for devices based on epitaxial graphene 4H-SiC, reaching a minimum detected concentration of 10 ppb [27].

In this work we present a wafer scalable nanopatterning method that combines a high edge length per unit area and the controlled generation of structural point defects by ion bombardment. Our approach is to use spherical block copolymer (s-BCP) lithography as a patterning mask, since this is robust and scalable compared to cylindrical block copolymers (c-BCP). We have fabricated graphene gas sensors from nanopatterned CVD graphene and exposed them to $NO_2$ in the 300 ppt to 100 ppb range. Our results present a drastic improvement of the response with respect to previous works on $NO_2$ graphene based gas sensors. The response was 7% at 300 ppt, which is the lowest measured concentration for $NO_2$ reported to date for a chemiresistive graphene sensor, except the work on UV illuminated sensors. Based on our Raman analysis of the defect density we suggest that the strong $NO_2$ response is due to the simultaneous contribution from edge adsorption sites and inter-hole point defects.

## 2 Results and Discussion

### 2.1 Nanopatterning using the PS-b-PMMA spherical block copolymer nanomask

CVD graphene transferred to oxidized silicon surfaces (for details of growth and transfer see the Experimental Section) was patterned into a nanomesh. The process sequence is described in Fig. 1a. Graphene is initially covered with a 15 nm thin layer of electron beam evaporated silicon dioxide ($SiO_x$). Afterwards, a 40-50 nm thick layer of polystyrene-b-poly methyl methacrylate (PS-b-PMMA) BCP is spin cast on top, without any pretreatment of the surface. For a 40-50nm s-BCP PS-b-PMMA film, the PMMA spheres segregate at the top of the film with a short-range hexagonal ordering (Fig. 1b and Fig. 1a-ii). The local order has been studied as a function of the annealing parameters and it was found to be highly robust with respect to variation of the main process parameters, maintaining the hexagonal order in the range 180 °C to 230 °C and in the 3h to 24h annealing time range.

The PMMA spheres are then removed using a flood UV exposure, followed by an acetic acid bath, leaving a nanoporous polystyrene thin film (Fig. 1a-iii). Reactive ion etching with an $O_2$/Ar plasma was used to etch the remaining polystyrene layer until the $SiO_x$ layer was reached. The directionality of the etching is ensured by the low pressure of the process and a low $O_2$/Ar ratio. As shown in Fig. 1b the $SiO_x$ layer is not reached for all the polystyrene nanopits at the same time (the white spots in the SEM images in Fig. 1b appears where the $SiO_x$ has been reached). This is to be expected, since the PMMA sphere diameter in this work varies between 10 to 50 nm. An optimization of the etching time is necessary in order to maximize the number of nanopits fully open, avoiding at the same time over-etching of the polystyrene nanomask. This process leaves a polystyrene holey layer of approximately 10 to 15 nm in thickness, which serves as a nanomask for the subsequent etching of the $SiO_x$ layer (Fig. 1b and Fig. 1a-iv).

When the polysterene nanomask is completed, the $SiO_x$ layer is etched using a $CHF_3$/$CF_4$ plasma (Fig. 1a-v). Afterwards, an oxygen plasma removes the residual BCP nanomask and etches the graphene through the holes in the SiOx mask. Finally, a 2 seconds dip in HF 5% is sufficient to remove the e-beam evaporated SiOx (Fig. 1a-vi). As shown in Fig. 2 the process time for the $SiO_x$ etching is critical in determining the morphology of the nanopatterned graphene. By changing this etching time, it is possible to tune the morphology of the nanopatterned graphene ranging from few sparse holes to a densely nanoporous, nearly discontinuous graphene layer. In Fig. 2c the fraction of removed graphene in percent and the corresponding edge length per unit area are shown for four different etching times. The same patterning process has been applied to exfoliated graphene and the comparison with CVD graphene can be seen in Fig. S-1 in the Electronic Supplementary Material (ESM). The size distribution of the holes is considerably broader as compared to typical results with c-BCPs, and for the longest etching time the short-range hexagonal order is lost. Fig. 2a presents a 4 $cm^2$ piece of CVD fully nanopatterned graphene.

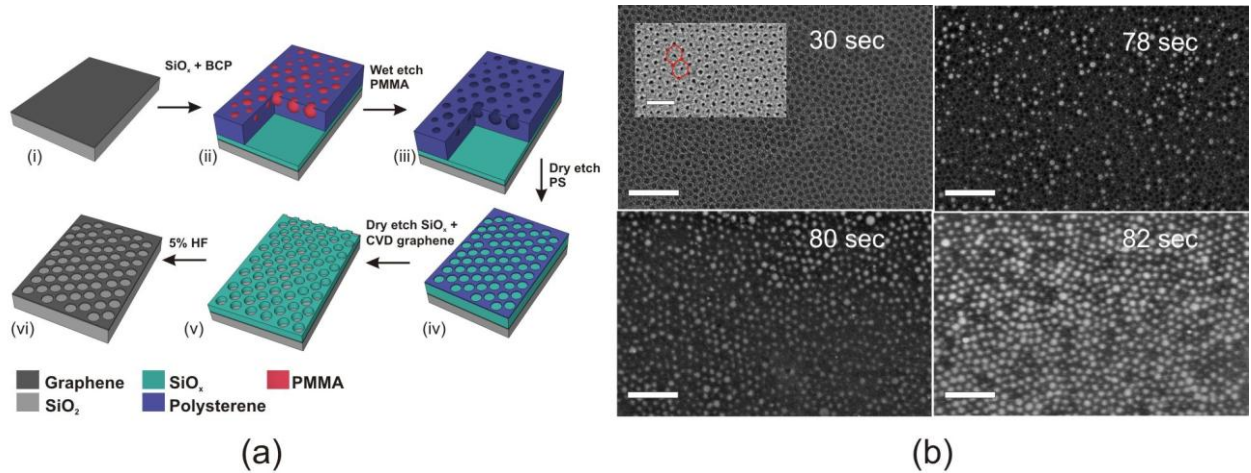

**Figure 1** (a) Microfabriaction sequence of the nanopatterning. (i) CVD graphene is transferred on silicon dioxide.(ii) It is covered with 15 nm of silicon oxide and the spherical block copolymer PS-b-PMMA is spin-cast on top and subsequently annealed. The cross-sectional view shows the inner geometry of the BCP. (ii) After an exposure to UV light the PMMA spheres are removed in acetic acid. (iv) An oxygen plasma is used to etch the polystyrene nanomask until the nanopits are all open. (v) A fluorine based plasma is used to etch the $SiO_x$ layer. An oxygen plasma removes the residual BCP and patterns the graphene. (vi) A 5% HF dip removes the remaining $SiO_x$. (b) SEM micrographs showing the top views of the polystyrene nanomask after different etching times in $O_2$/Ar plasma. The inset shows a SEM micrograph of the top view of the PS-b-PMMA BCP before etching. After 30 seconds no nanopits have reached the $SiO_x$ layer. After 78 seconds of etching, approximately ~50% of the pits are have reached the SiOx layer. After 80 seconds of etching, ~90% of the pits are cleared. After 82 seconds of etching, ~95% of the pits reach the $SiO_x$ layer and some of the holes are merged. The scale bars are 300 nm (200 nm in the insert)

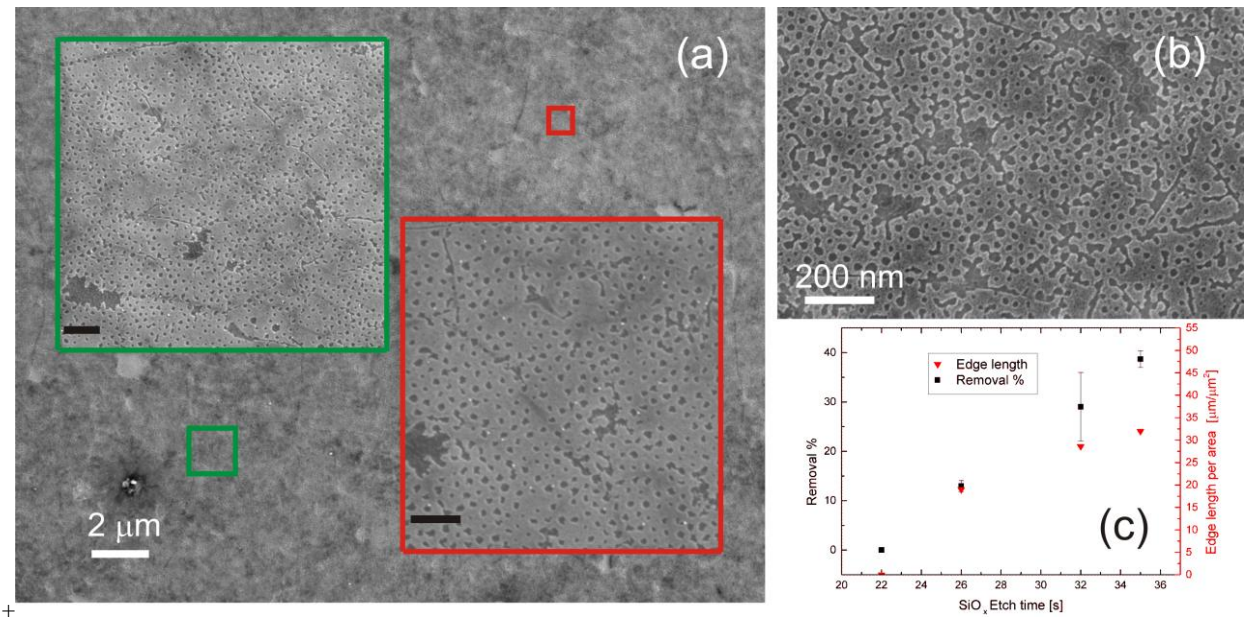

**Figure 2** (a) SEM micrographs of CVD nanopatterned graphene. a) Large view of CVD graphene when 26 seconds etching of the SiOx protective mask is used. The insets show the nanopattern. (b) Nanopattern when 32 seconds etching of the SiOx protective mask is used. The scale bars are 200 nm for insets.(c) Graphene removal percentage and edge length per area ratio for nanopatterned CVD graphene. For 22 seconds of SiOx etching the pattern is not transferred. After 26 seconds 13% is removed with an edge length per unit area of 19 μm/μm2. For 32 seconds the 28% is removed with an edge length per unit area of 28 μm/μm2. For 35 seconds 38% is removed with an edge length per unit are of 32 μm/μm2.

Typically, a random copolymer, which contains the same polymeric components as the BCP, and thus exhibits the average surface energy of the two polymers, is used to ensure the vertical alignment of the cylindrical block copolymer on top [23]. Spherical BCP are always correctly oriented instead, so no fine-tuning of the surface energy is needed as long as the polymer wets the surface – this makes processing and formation of homogeneous spherical BCP lithography mask intrinsically simpler than for cylindrical BCP. While the s-BCP structures are more disordered than the c-BCP and colloidal nanosphere lithography, no evidence of the long-range order to play any role in the sensing properties has been reported. The edge length per unit area achieved with our nanopatterning technique is up to 32 $\mu m/\mu m^2$ (Fig. 2c) and compared to previous works on graphene nanomesh for gas sensing, the edge length per unit area and thus the estimated density of edge adsorption sites for gas molecules is a factor of 2 higher [18].

So far very few works have reported a nanopatterned area larger than 1 $cm^2$ using BCPs or any other technique [18,24,28]. This is because it is difficult to achieve a uniform holey mask across $cm^2$ large areas. The morphology of our BCP film after annealing as well as after etching is highly uniform across an entire 4" inch wafer (see Fig. S-2 in the Electronic Supplementary Material (ESM)), but a larger nanopatterned area could not be achieved due to space limitations in the e-beam deposition system used to deposit $SiO_x$. Finally, several factors can affect the uniformity at different length scales, in particular microscale particles that introduce variation in the BCP layer thickness, and the uniformity of the etch rate of the RIE machine. These factors, however, would have a similar impact on other lithographic techniques.

## 2.2. Electrical and gas measurements

After nanopatterning, the three different CVD graphene films with 13%, 28% and 38% of the total area removed was transferred using a wedging process onto a silicon substrate with 300 nm thermal oxide and lithographically defined 4 μm wide electrical contacts (5 nm Cr/45 nm Au), as shown in Fig. 3a [29]. A device is shown in Fig. S-3. In order to characterize the transport characteristics and the gas response of our nanopatterned graphene a Linkam LS600P enclosed custom probe station was used. Gas measurements were performed in an atmosphere of synthetic air with a small (0.3 bar) under-pressure applied to the chamber. The system allowed for concurrent measurements of two devices, so that a nanopatterned sample could be compared to a non-patterned sample simultaneously. The electrical resistance of the graphene devices was measured as a function of gas concentration and temperature. All the devices showed p-doping at zero-gate bias, most probably a consequence of the transfer process [30]. Therefore, lowering in the Fermi level, due to the charge transfer to $NO_2$ molecules, corresponds to a decrease in the total resistance. It is important to notice that the conductance versus gate voltage characteristics of the CVD nanopatterned graphene maintained the typical graphene like shape, showing p-doping and lower conductance values with respect to non-patterned CVD graphene (see Fig. S-4 in the Electronic Supplementary Material (ESM)). All the measurements were performed following the same experimental sequence. The graphene devices were initially annealed in nitrogen atmosphere at 225 °C for two hours to remove gas and water molecules adsorbed on the graphene. After stabilizing the temperature, $NO_2$ was permitted into the chamber for 120 seconds followed by a recovery period of 240 seconds, with the exception of sub-ppb gas concentrations, where the active measurement time was increased from 120 seconds to 20 minutes. Three gas injections were done for each set of conditions, followed by annealing at 225 °C for five minutes in order to remove residual gas molecules before the next experiment.

Direct comparison of unpatterned to nanopatterned CVD graphene showed a significantly higher response to $NO_2$ for nanopatterned graphene. Fig. 3b shows a typical response to 100 ppb $NO_2$. At t = 60 seconds, 100 ppb $NO_2$ is allowed into the gas chamber for 120 seconds followed by 240 seconds recovery phase. The black line shows a non-patterned device response, which exhibited a -1.2% change (after first gas injection), which is considerably less than the -7.5% for the nanopatterned graphene. In order to quantitatively compare the gas responses of our graphene devices the relative changes 2 minutes after each injection will be used in the subsequent analysis.

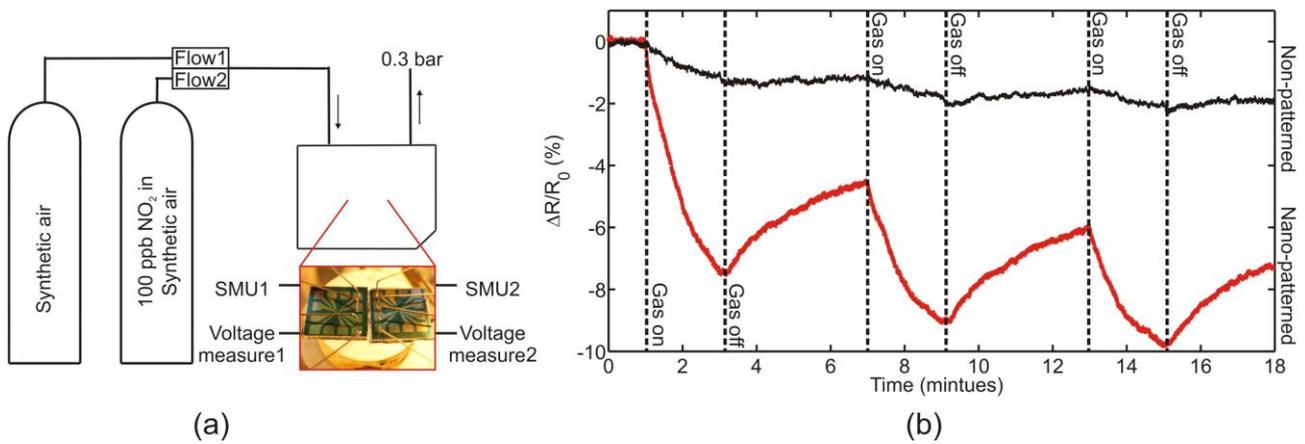

**Figure 3** (a) The set-up used: a CVD nanopatterned device and a CVD non-nanopatterned device are placed in the chamber at the same time and measured using two voltage sources and two source measure units (SMU). (b) Typical response to 100 ppb $NO_2$ with the response of nanopatterned graphene (38% removed) shown in red and non-patterned shown in black.

**Table 1** Gas Response data after the first gas injection of the non-patterned and nanopatterned devices.

| Sample | Sensitivity 1 ppb | Sensitivity 10 ppb | RMS noise | SNR at 1 ppb | SNR at 10 ppb |
|---|---|---|---|---|---|
| Non-patterned | 0.15% | 0.36% | 0.2% | 1< | 2 |
| 13% | 1.8% | 4.5% | 0.57% | 3 | 8 |
| 28% | 2.6% | 6.6% | 0.09% | 29 | 73 |
| 28% | 4.4% | 10% | 0.26% | 17 | 43 |
| 38% | 6.2% | 7.5% | 0.39% | 16 | 19 |

Fig. 4a shows the room-temperature response 2 minutes after the first injection of various samples to $NO_2$ concentrations of 1 ppb, 10 ppb and 100 ppb. For all concentrations, the nanopatterned graphene response is significantly larger than that of the non-patterned graphene and the expected correlation between concentration and the device response was observed, where an increase in the concentration corresponds to an increased response. The sensitivity at 10 ppb and 1 ppb, the level of noise and the signal to noise ratios (SNR) for the devices are summarized in Table 1. An increase of the SNR up to of a factor 37 is observed with respect to non-patterned CVD graphene.

In Fig. 4b-c the sensor responses after all the three injections at 1 ppb and 10 ppb are presented at room temperature and 175°C respectively. In the Electronic Supplementary Material (ESM) the data for 100 ppb at 175 °C are disclosed. As can be expected the sensor response is decreasing after each gas injection, even if the nanopatterned sensors remain significantly more sensitive than the non-nanopatterned. The decrease is due to the saturation of the adsorption sites, which is not fully recovered in the 4 minutes of desorption time in between the gas injections. The response decrease is more pronounced for the nanopatterned devices, indicating that the binding energy of the adsorption sites is different. This is coherent with the fact that the nanopatterned devices have a high density of defects which are known to have higher binding energies than the graphene basal plane [14-16]. The difference in binding energy of $NO_2$ molecules is supported by the fact that the decrease of gas sensitivity is reduced when the sensor is operated at 175°C, allowing for a faster desorption between the gas injections.

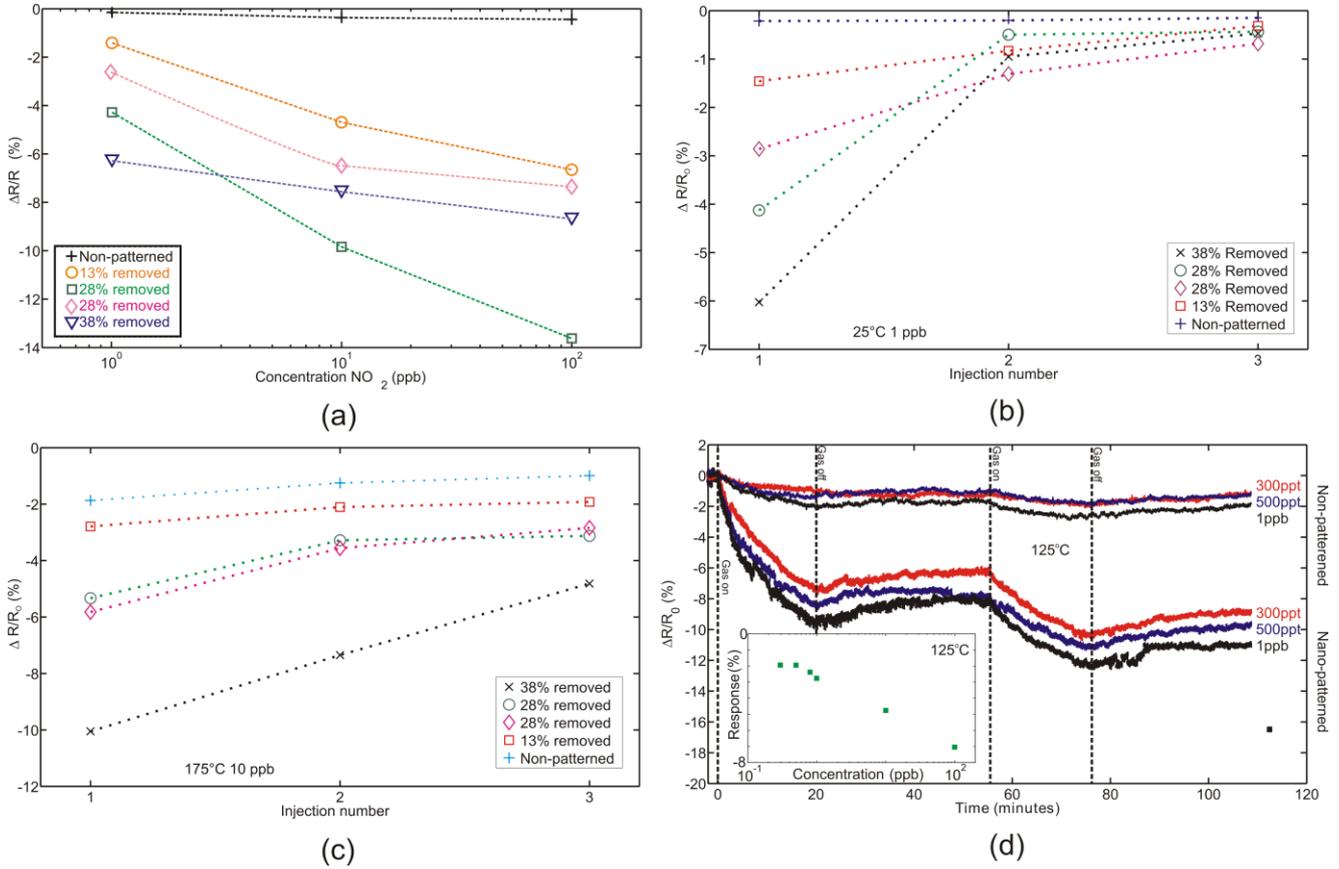

Figure 4. a) Response of all samples at 25°C, 2 minutes after the first injection of 1, 10, 100 ppb NO2. b) Response of all samples after each injection for 1 ppb at 25 °C . c) Response of all samples at 10 ppb at 175 °C. d) Response of non-patterned versus nanopatterned sample (28% removed) at 125°C for sub-ppb concentrations with 20 minute NO2 exposure intervals. Inset: Response for a 28% removed sample at 125°C over at different concentrations.

In order to test the minimum detectable concentration the NO$_2$ concentration was lowered below 1 ppb, by increasing the dilution. Fig. 4d shows a clear sub-ppb response for a nanopatterned graphene device with 28% removal. The larger absolute responses compared to Fig. 4a are due to the prolonged active measurement period of 20 minutes. Even at the lowest dilution, which could be reached accurately (300 ppt) the sensor reacts with change of approximately -7% (2% after 2 minutes). The signal to noise ratio in this case is 26, suggesting that much lower concentrations can be detected.

While there is a clear tendency of response enhancement from nanopatterning, and sub 300 ppt detected concentrations have been demonstrated, the devices show a too large variation in the NO$_2$ response to draw any defined conclusion on the quantitative relationship between removal percentage and gas response. However, it should be noticed that the 13% removed device is less sensitive than 28% removed and 38% removed devices in almost all conditions of temperature and concentration. This is consistent with the notion that shorter edge length per area gives less adsorption sites for the NO$_2$ molecules and that shorter etching time generates fewer inter-hole point defects. By comparing the 2 minutes response values in Fig. 4 with the most sensitive sensors from other works, our responses are much larger and the response times are dramatically faster. For instance, the CVD nanomesh graphene sensor reports a response similar to Fig. 4, however, with a 15 minute response time and for a 1000 times higher concentration [18]. This drastic difference cannot be ascribed only to the increased edge length, because, as mentioned in section 2.1, the edge length per unit area achieved with our nanopatterning process is just twice that achieved by the nanomeshed chemisensors [18]. This indicates that the intimate nature of the CVD nanopatterned graphene used is very different. In section 3 this difference is attributed to the high density of inter-hole point defects. Comparing with the work of Yavari et al., the response at 100 ppb of their CVD based sensors and our non-patterned CVD sensors are similar after 2 minutes (~-1%). However, our nanopatterned sensors reach much larger responses (up to -13.8%) in just 2 minutes at 100 ppb, giving more than one order of magnitude improvement in the response [8]. Comparing with the ozone treated CVD graphene, which shows a response of --0.5% after 20 minutes at 200 ppb, we have a response up to 26 times larger in only two minutes at 100 ppb [17]. Novikov and coauthors showed a -7% change upon exposure to 10 ppb after 20 minutes, whereas our sensors reached the same relative change for a concentration 33 times lower [27]. Finally, the signal-to-noise ratios in Table 1 and data in Fig. 4 suggest that much smaller concentrations could be detected and detection limits in the tens of ppt range are achievable with a basic chemiresistive graphene sensor. In literature, it is common to extrapolate the detection limit of a graphene gas sensor, based on the supposition of a linear relation between sensor response and concentration of $NO_2$ [17,18,25,31]. The detection limit may then be inferred as the RMS noise of the measurement divided by the slope of the linear curve response versus concentration. This approach relies on the very strong assumption that the linear response can be extrapolated outside the measured range by several orders of concentration magnitude, since the measurements are done at much higher concentrations with respect to the calculated detection limit. Our data in Fig. 4 in the 100 ppb to 300 ppt range contradict this behavior showing a strong non-linearity (note the logarithmic scale). To the best of our knowledge, the data presented in this work are the first data presented for sub 10 ppb concentrations for not UV illuminated graphene based sensors.

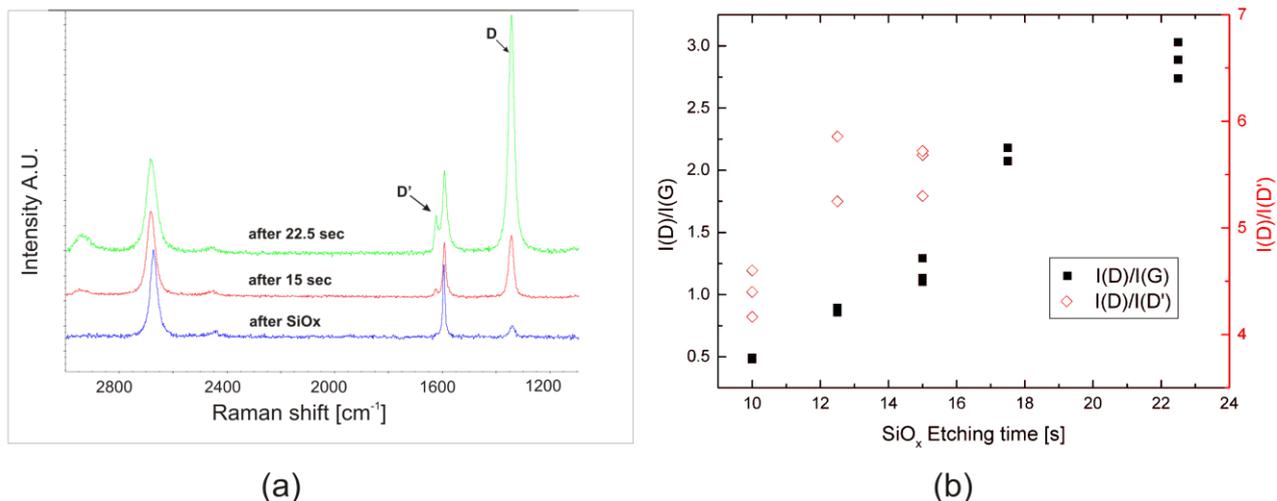

Figure 5. (a) Graphene Spectra after SiOx deposition and etching. (b) I(D)/I(G) and I(D)/I(D') ratios measured on SiO2/exfoliated graphene/SiOx samples as a function of the etching time using the fluorine based plasma recipe in the RIE chamber. The I(D)/I(G) ratio increases for longer etching times. Three spectra were taken per each sample.

## 2.3 Defect analysis by Raman spectroscopy

In order to understand why our samples have so high sensitivity, we investigated by Raman spectroscopy whether the nanopatterning process in addition to creating nanosize holes in the graphene also generates point defects in the graphene between the nanoholes. To perform this analysis we used non-nanopatterned SiO2/exfoliated graphene/SiOx structures. In this case the single layer exfoliated graphene flakes on thermal SiO2, with no visible D-peak and an I(2D)/I(G) ratio larger than 1 (laser wavelength = 532 nm) were covered by 15 nm SiOx as described before for the nanopatterned samples. The absence of nanopatterning or initial D-peak in these test structures permitted to attribute a possible D peak evolution to point defects only and not at the edges as in the case of nanopatterned samples. The Raman spectra were acquired leaving the remaining SiOx on top of the exfoliated graphene, since we checked that there is no measurable difference in the spectra after removing the SiOx layer in BHF. After the SiOx deposition step an I(D)/I(G) ratio of 0.18 was found, indicating that the graphene is slightly damaged by the $SiO_x$ deposition (see Fig. 5a). The samples were then etched with the same reactive ion etching recipe used for nanopatterning of the $SiO_x$ layer (see Fig. 1a-v). The etching time was increased in steps until the point where the $SiO_x$ layer was entirely removed. Ellipsometer measurements showed 5 nm retained after 12.5 seconds, 2 nm retained after 17.5 seconds and the SiOx layer is etched away after 22.5 seconds. We recorded the evolution of two defect activated peaks, D peak at ~1340 cm$^{-1}$ and the D' peak at ~1620 cm$^{-1}$, as a function of the $SiO_x$ etching time. The evolution of the Raman spectra is shown in Fig. 5a, whereas the intensity of the I(D)/I(G) ratio as function of the etching time is presented in Fig. 5b. The I(D)/I(G) ratio increases for longer etching times, until reaching a value of 2.9 (see Fig. 5b).

The evolution of the I(D)/I(G) ratio indicates that the point defect density in the exfoliated test graphene increases for longer RIE etching times, despite the presence of the initial 15 nm SiOx on top of the graphene. This trend can be explained by the energetic ions accelerated by the DC voltage inside the RIE chamber towards the SiOx layer being capable of penetrating the progressively thinner evaporated silicon oxide and damaging the graphene lattice underneath. Following the method described by Cançado et al., it is possible to calculate the point defect density expressed by the characteristic distance between defects $L_D$ [32]. In their work they related the I(D)/I(G) ratio with $L_D$ deriving the formula $L_D^2 = 1.8 \times 10^{-9} \times \lambda^4 (I(D)/I(G))^{-1}$, where $\lambda$ is the laser wavelength. The obtained $L_D$ values for our samples are 17 nm, 13 nm, 11 nm, ~7 nm, ~5 nm, for etching times of 10 s, 12.5 s, 15 s, 17.5 s, 20.5 s and 22.5s, respectively. Since the formula used is valid only for $L_D$ > 10 nm, the last two values are read from Fig. 3 in ref [32]. Considering that the $SiO_x$ layer was entirely removed from 17.5 to 22.5 seconds etching time, we conclude that our nanopatterned graphene must have a defect characteristic length close to $L_D$ = 7 nm, corresponding to a density of one defect per 265 carbon atoms. Therefore, comparing the dimensions of the nanopattern in Fig. 2, where the typical neck width between the nanoholes is significantly larger than 7 nm, we can conclude that defects are very dense compared to the BCP created nanopattern. Since a strong enhancement in the gas response has been measured for graphene with much lower defect densities, we expect that the inter-hole point defects originating from the etching process are playing an important role in the high gas response of our sensors [16]. Finally, the evolution of the I(D)/I(G) ratio can elucidate the degree of lattice disorder of the samples. According to the amorphization trajectory for graphene described in ref [32], a value of the I(D)/I(G) ratio of 3 (green laser) is the delimitation value between end of stage I and beginning of stage II. The nanopatterned CVD graphene investigated here should then be at the end of stage I, with mainly sp2 hybridization and nanocrystalline lattice order.

We note that in the previous works concerning graphene nanomesh for electronic application, a 10 nm thin SiOx layer was used on top of graphene [19,20]. Neither of these discussed the possibility of the graphene below the mask to be damaged as well between the holes. Based on our results, we anticipate that co-generation of inter-hole defects may be a general issue for block copolymer lithography of graphene, which will become progressively more serious as neck width and thus mask thickness decrease.

Recent works by Eckmann et al. [33,34] pointed out that the I(D)/I(G) intensity ratio alone is not sufficient to determine the nature of the defect, but that the I(D)/I(D') ratio behaves distinctly differently depending on the type of defect. To investigate the nature of the point defects between the holes generated during the silicon oxide etching plasma, we analyzed the I(D)/I(D') ratio in our exfoliated samples (see Fig. 5b). In our case we measured an I(D)/I(D') ratio evolving from ~4.5 to ~5.5 with the etching time.

Eckmann and coauthors reported that sp3 sites, vacancies, substitutional Boron atoms and grain boundaries defects show an I(D)/I(D') ratio of ~13, ~7,~9,~3.5 respectively [33,34]. These ratio values were measured in samples dominated by a single type of defect with the specific ratio I(D)/I(D') remaining constant for a wide range of defect densities in the regime $L_D$ > 10 nm . Since our I(D)/I(D') ratio values evolve from ~4.5 to ~5.5 (for $L_D \geqslant$ 11 nm), this methodology suggests that the ion-bombardment through the SiOx layer is generating a mixture of grain boundary-like defects and vacancies and that the relative generation rate of these two types of defects changes with the energy of the ions reaching the graphene, due to the gradual thinning of the SiOx layer. The creation of boundary-like defects is particularly interesting because, it has been shown that Stone-Wales defects (typical grain boundaries structures) are the most efficient type of defect in binding $NO_2$ molecules [14,16]. However, the creation of boundary like defects has to be confirmed by further experiments, since ion bombardment is generally considered a source of vacancy like defects [33]. Moreover, a higher concentration of grain boundaries has been used to justify a higher sensitivity to $NO_2$ for different types of CVD graphene [18].

## 3. Conclusions

We report readily scalable nanopatterning technique for $cm^2$ scale CVD graphene. The fabricated $NO_2$ gas sensors using the nanopatterned CVD graphene demonstrate unprecedented conductivity response in the sub 100 ppb range of $NO_2$. Resistance changes up to -6.7% and -10% have been measured after just 2 minutes exposure at 1 pbb and 10 ppb concentrations. Moreover, the high measured signal-to-noise ratio at a concentration of 300 ppt, indicates that tens of ppt resolution is within reach. Based on the Raman analysis of the defect density, we suggest that the observed sensitivity may not just be accounted for by the large edge length per area of the dense nanopatterned structures, but also by the generation of point defects between the nanoholes despite presence of a thin SiOx etch mask. Such defect generation under the etch mask could have implications for the application of BCP lithography for other electronic applications, as defects in nearly all other such applications will be detrimental to the performance. Areas as large as 4 $cm^2$ were patterned without any substrate surface treatment, and were in our case limited by the available physical sample space in one of the necessary processing steps. This work shows that ultrasensitive $NO_2$ gas sensors can be fabricated with a robust, scalable nanopatterning technique.

## 4. Experimental methods

CVD graphene was grown on a copper foil in an Aixtron Black Magic cold wall CVD system following a typical, common recipe [7]. The graphene was then transferred on oxidized silicon wafers by etching the metal growth substrate with ammonium persulfate [28]. The block copolymer used is PS-b-PMMA (195K-20K) purchased from Polymer Source. The block copolymer annealing carried out at 230 °C for 12 hours in low vacuum. In order to remove the PMMA spheres the samples were exposed to UV light (I-line) with an intensity of 7 W/$cm^2$ for 45 seconds, and subsequently immersed in acetic acid for 15 minutes and then rinsed with DI water. A plasma containing oxygen and argon (5 sccm and 45 sccm, respectively) at a pressure of 10 mTorr and 30W of power was used in a Reactive Ion Etching (RIE) machine to etch the remaining nanoporous polystyrene layer in order to form the nanomask. The same RIE machine was used to etch the SiOx layer. In this case the plasma contained $CHF_3$ and $CF_4$ (26 sccm and 14 sccm, respectively) at a pressure of 100 mTorr and a power of 60 W. The final recipe used to remove the residual polystyrene contains oxygen and argon (45 sccm and 5 sccm, respectively) and the machine was operated at 10 mTorr and 30 W of power for 30 seconds. The remaining electron beam evaporated silicon dioxide was removed with a 2 seconds dip in 5% HF, followed by a rinse in DI water and isopropanol. In order to characterize the samples, a Supra 40VP Zeiss SEM microscope and ThermoFisher DXR Raman system was used with a 532 nm wavelength at 0.8 mW power. Keithley 2400 source measure units (SMU) were used to supply voltage and measure current within a customized Linkam LS600P environmental measurement chamber. Device temperature could be controlled between 123 K and 423 K ± 0.1 K. Keithley 2000 voltmeters were used to measure four-point voltages, while the current was sourced by a Keithley 2400 Sourcemeter. Source and measurement instruments, device temperature, gas flow and recording of data were controlled via LabView. The 100 ppb $NO_2$ in synthetic air bottle was diluted in 99.999 % synthetic air using Tylan flow controllers connected in parallel.

# Acknowledgements

The authors wish to thank the Center for Nanopatterned Graphene and the European Project NMP-FP7 Grafol for the financial support. The Center for Nanostructured Graphene (CNG) is sponsored by the Danish National Research Foundation, Project DNRF58. The authors would like to thank Dr. Tim J. Booth for the fruitful discussion and DTU-Danchip personnel for the help during the microfabrication.

**Electronic Supplementary Material**:
A comparison between nanopatterned CVD graphene and micro exfoliated nanopatterned graphene is presented, as well as a full wafer s-BCP sample. A current-gate voltage characteristic of nanopatterned CVD graphene is shown, along with and optical image of one of the devices. The response of all devices after each injection for 100 ppb $NO_2$ is also presented.

# Electronic Supplementary Material

# Large-Area Nanopatterned Graphene For Ultrasensitive Gas Sensing

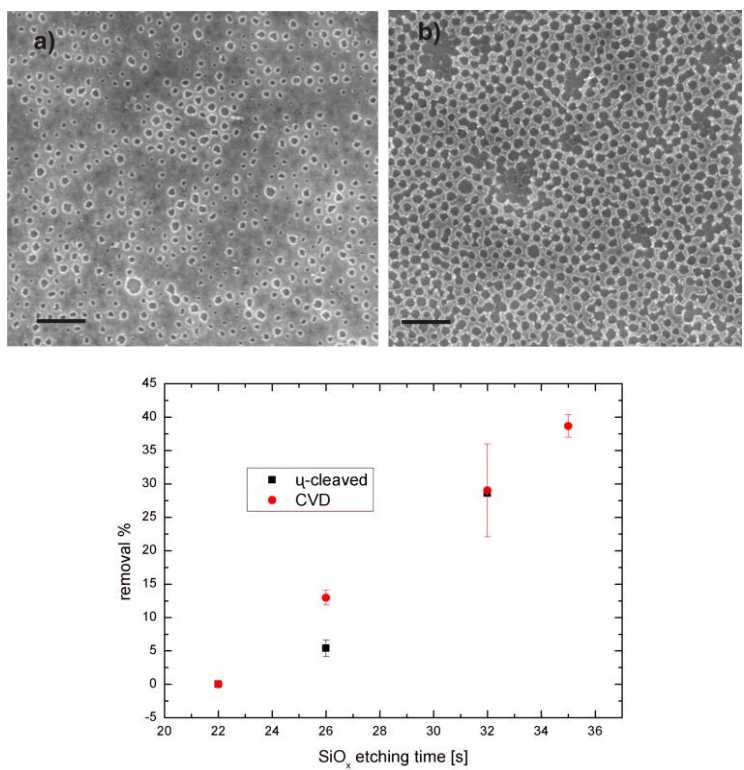

**Figure S-1** Nanopatterned graphene using spherical block copolymer was realized also for micro-cleaved grapehene. Single layers were identified and patterned using the same recipe described for CVD grapehene. The pattern removal percentages and the morphology of the patterns are similar for the two cases. a) 26 seconds for the SiOx etching, b) 32 seconds for the SiOx etching. The scale bars are 100 nm.

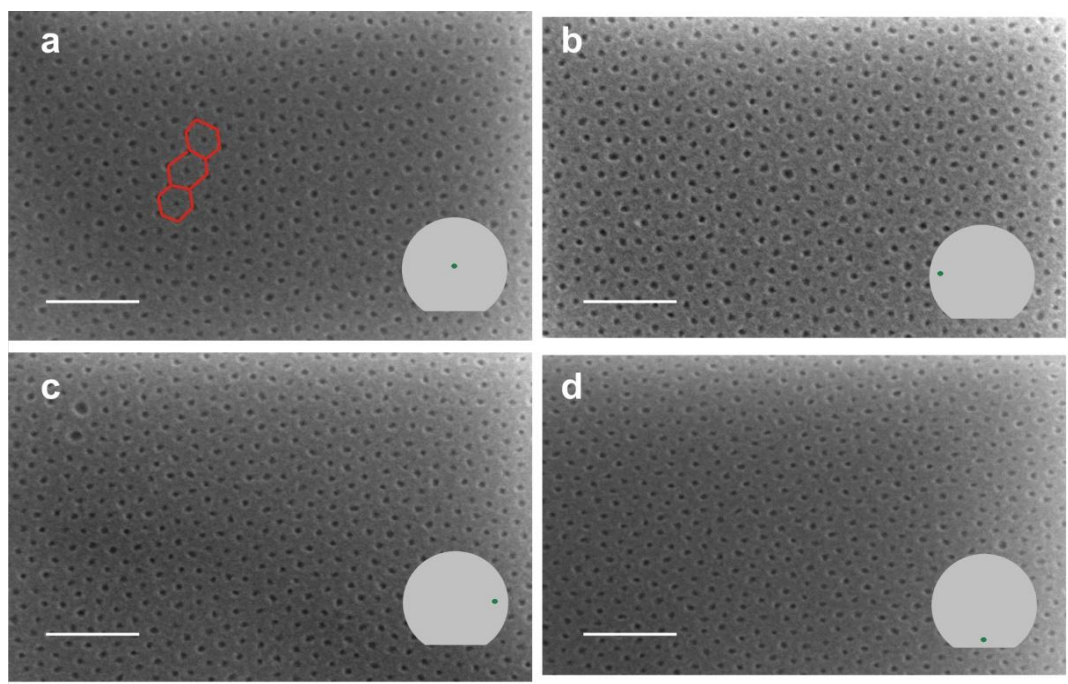

**Figure S-2** Scanning Electron Microscope (SEM) micrographs of 4 different positions on a 4 inches wafer cover with the spherical block copolymer. After the annealing the PMMA spheres are organized in a hexagonal pattern with a short range order across the entire wafer. (a) center of the wafer, (b) left, (c) right, (d) bottom. The green dots indicate the position where the image was taken. The scale bars are 200 nm.

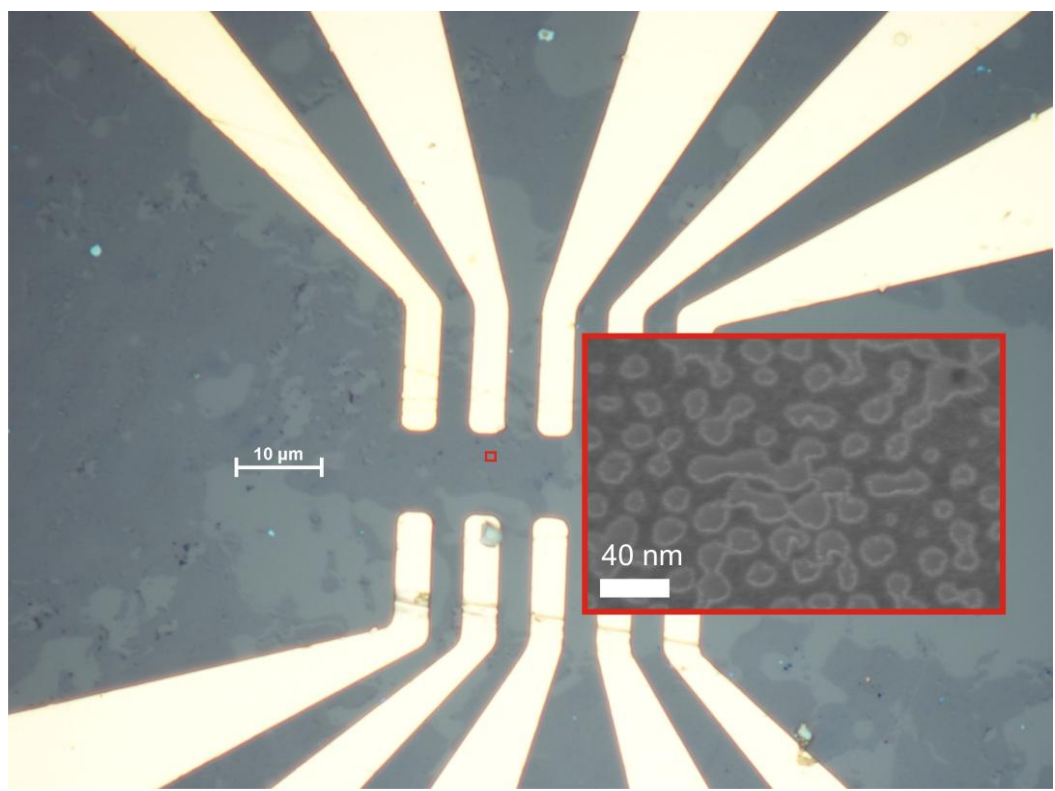

**Figure S-3** Optical image of one of the device. CVD nanopatterned graphene is place on top of Gold electrodes and the polymer used for the wedged transfer is dissolved in ethyl acetate. The insert present a SEM image of the nanopatterned graphene. The electrical measurements are done using four electrodes.

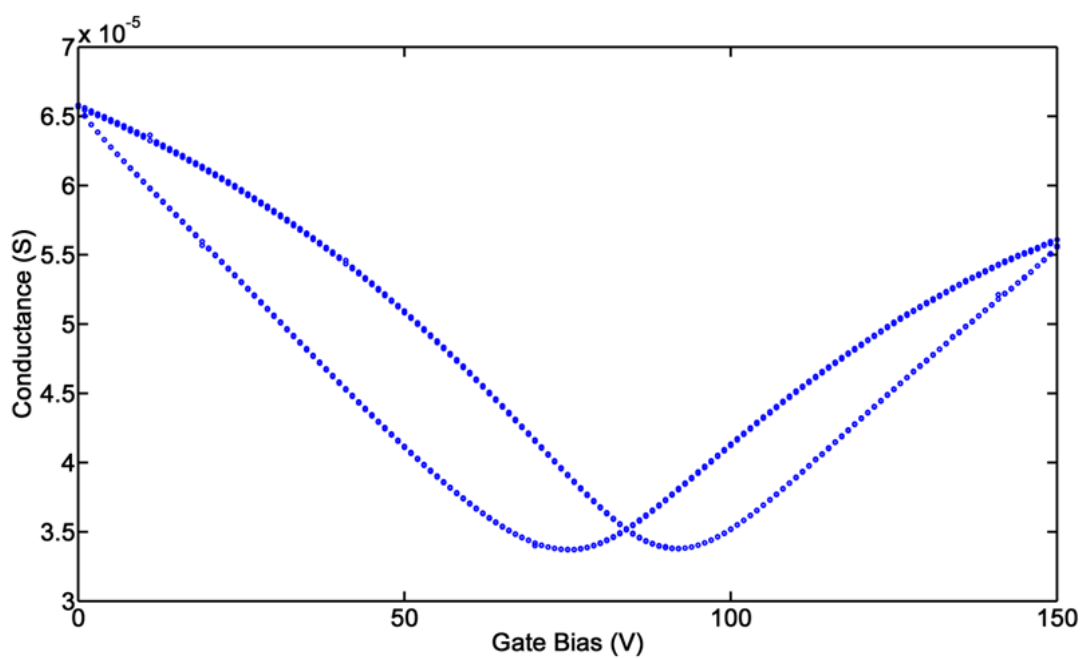

**Figure S-4** Typical condunctance/Gate Voltage characteristic of the 28% removed nanopatterned graphene. All our devices showed similar charateristics, indicating that our graphene has low condunctance and it is p-doped. It should be noticed that a 300 nm oxide has been used.

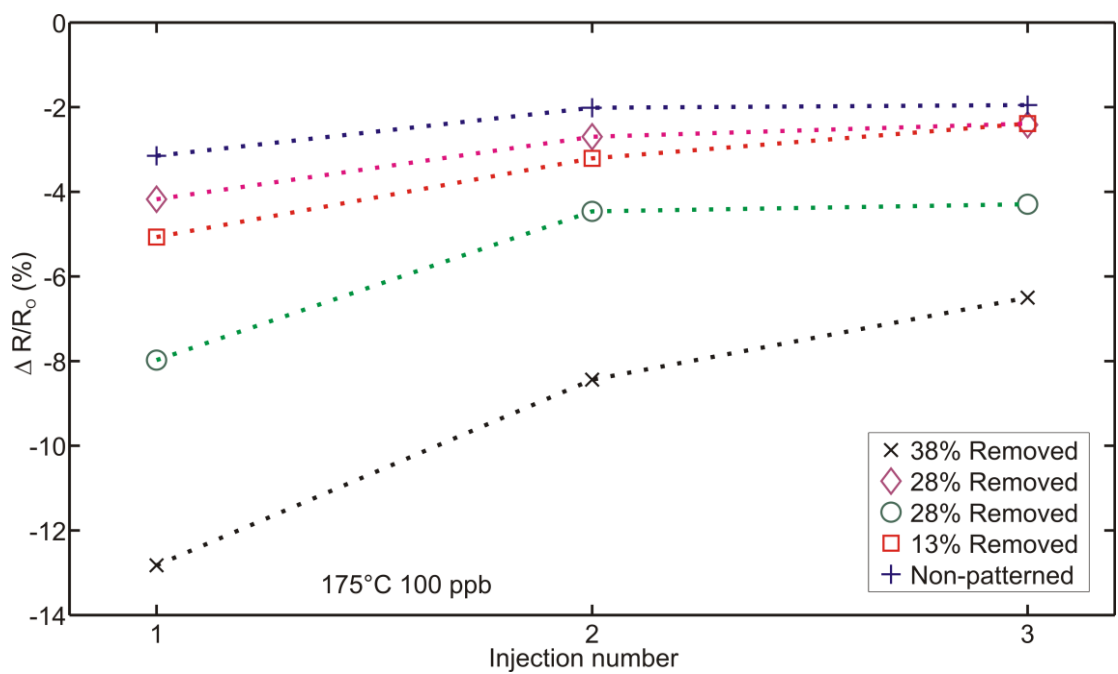

**Figure S-5** Response of all samples after each injection of 100 ppb of $NO_2$ at 175 $^o$C. The response of all samples is reduced after each injection, but the nanopatterned devices show higher response.

Address correspondence to Alberto Cagliani, alberto.cagliani@nanotech.dtu.dk; Peter Bøggild, peter.bøggild@nanotech.dtu.dk;